\documentstyle[prd,eqsecnum,preprint,tighten,aps]{revtex}
 \newcommand{\R}{\mathbf{R}}

\begin{document}
 \renewcommand{\theequation}{\thesection.\arabic{equation}}

 \title{Topological Constraints on the Charge Distributions for the Thomson Problem}
 \author{Alfredo Iorio$^{a,b}$\thanks{E-mail: iorio@lns.mit.edu},
 Siddhartha Sen$^{c,d}$\thanks{E-mail: sen@maths.tcd.ie, tcss@mahendra.iacs.res.in}}
 \address{$^a$ Department of Physics, Brown University \\
 182, Hope Street, Providence RI 02912 - USA \quad {\rm and}}
 \address{$^b$ Department of Physics ``E.R. Caianiello'', Salerno University and I.N.F.N. \\
 Via Salvador Allende, 84081 Baronissi (SA) - Italy}
 \address{$^c$ School of Mathematics, Trinity College Dublin, Dublin
 2 - Ireland \quad {\rm and}}
 \address{$^d$ I.A.C.S., Jadavpur, Calcutta 700032 - India}

 \date{\today}

\maketitle

\vskip .5cm

\begin{abstract}
The method of Morse theory is used to analyze the distributions of
unit charges interacting through a repulsive force and constrained
to move on the surface of a sphere -- the Thomson problem. We find
that, due to topological reasons, the system may organize itself
in the form of pentagonal structures. This gives a qualitative
account for the interesting ``pentagonal buttons'' discovered in
recent numerical work.
\end{abstract}

\vfill

\noindent {\it Dedicated to Rafael Sorkin on his 60th birthday}

\bigskip
\bigskip

%\noindent PACS No.: ***

%\noindent Keyword(s): ***

\noindent Brown-HET-1463

\noindent cond-mat/0603044

\newpage

%%%
\section{Introduction}
%%%

The Thomson problem consists in determining the minimum energy
configurations for a given collection of $N$ unit like-sign
electric charges constrained to move on the surface of a
two-sphere of radius $R$ \cite{Thomson}. The Coulomb energy of
such a system is
\begin{equation}\label{ethom}
  E_C \sim \sum_{i < j} \frac{1}{\left| \vec{r}_i - \vec{r}_j
  \right|} \;,
\end{equation}
where $\vec{r}_i$ is the radial coordinate of the $i^{\rm th}$
charge on the sphere. The actual values of the electric charge and
of the dielectric constant of the medium are inessential for the
qualitative considerations we are going to make throughout the
paper.

This is an old and largely unsolved problem that, in its
generalized version -- i.e. for more general repulsive potentials
as well as for topological defects rather than unit electric
charges -- finds applications ranging from superconductivity to
biology.

In recent years, triggered by the seminal paper of reference
\cite{Bowick1}, there has been a lot of interesting theoretical,
numerical and experimental work on the generalized Thomson problem
\cite{Bowick2}. In \cite{Bowick1} the authors deal with
disclination defects constrained to move on a sphere. They show
how an \textit{effective} defect model, rather than the analysis
of the actual elementary charges interaction, proves to be
extremely reliable to describe (by numerical means) the ground
state configurations in terms of the ratio of the defect core
energies to the Young modulus. This way they provide interesting
empirical {\it solutions} to the Thomson problem, presenting
patterns the defects form on the surface of the sphere at the
various energy thresholds.

We would like to understand the patterns formed the way described
in \cite{Bowick1}, i.e. {\it why} these charges (defects) arrange
themselves on the surface of the sphere following precise symmetry
prescriptions. We would expect the distribution to have the
property that each charge has the same environment. If such an
assumption is made the charges should be distributed in a way that
represents a \textit{tiling} of the surface of the sphere.

The general problem of tiling a genus $g$ surface, say $\Sigma_g$,
can be solved in a simple way. Suppose we want to tile $\Sigma_g$
with regular $p$-sided polygons ($p$-gons) assembled in such a way
that each vertex is shared by (is a common vertex for) 3 $p$-gons,
and each edge is shared by (is a common edge for) 2 $p$-gons. If
$k_p$ is the number of $p$-gons used\footnote{There are, for
instance, $k_5$ 5-gons (pentagons), $k_6$ 6-gons (hexagons), $k_7$
7-gons (heptagons), etc.} to tile $\Sigma_g$, the resulting
\textit{polyhedron} $P$ has $V_P = 1/3 \sum_k k_p p$ vertices,
$E_P = 1/2 \sum_k k_p p$ edges, and $F_P = \sum_k k_p$ faces,
giving for the Euler characteristic $\chi (\Sigma_g) = V_P - E_P +
F_P$, the following expression
\begin{equation}\label{ngons}
  \sum_k k_p (6 - p) = 6 \chi (\Sigma_g) = 6 (2 - 2 g) \;.
\end{equation}
Generalization to the case of vertices shared by not always three
$p$-gons can be easily constructed.

For the sphere $g=0$, and if only \textit{hexagons} and
\textit{pentagons} are used, there can be an arbitrary number of
hexagons, but there must be exactly 12 pentagons. This follows
from Eq.(\ref{ngons})
\begin{equation}\label{sphere}
    \sum_k k_p (6 - p) = k_5 \; (6 - 5)  + k_6 \; (6 - 6) = 12
\end{equation}
hence $k_5 = 12$ and $k_6$ is arbitrary. If one also uses {\it
heptagons} for the tiling of the sphere, then one has the
interesting result
\begin{equation}\label{heptagons}
k_5 - k_7 = 12 \;.
\end{equation}
Thus, one can start off by tiling the sphere with an arbitrary
number of hexagons and exactly twelve pentagons. Then one can go
on by adding an arbitrary number of {\it pairs} pentagon-heptagon,
but not a pentagon or a heptagon separately. Let us notice {\it en
passant} that other ways of tiling the sphere are, of course,
possible. For instance, if only equilateral triangles ($3$-gons)
are used, then from Eq.(\ref{ngons}) follows
\begin{equation}\label{tetrahedron}
k_3 \; 3 = 12 \quad {\rm or} \quad k_3 = 4 \;,
\end{equation}
that is the {\it tetrahedron}.

We are particularly interested in the 5-gon--6-gon--7-gon tiling
because the authors of \cite{Bowick1} consider the pentagons and
heptagons as disclination defects in a sea of hexagons. They
construct an effective theory for $E_C$ in (\ref{ethom}) in terms
of interactions between the defects, the background charge
distribution of hexagons merely providing the value of the
effective Young modulus. They prove that for a sphere of large
radius, distorting the curvature by the introduction of defects is
not excessively expensive in terms of bending energy. A 5-gon
defect makes the local curvature negative, while a 7-gon makes it
positive. For a large spherical system, therefore, the
introduction of a 5-gon--7-gon pair is an energetically reasonable
way of joining a group of 12 charges. A calculation had to be
carried out to see if such a charge configuration was
energetically favored as compared to 12 charges organized on two
additional hexagons.

They find that when the total number $N$ of charges on the sphere
exceeds a certain critical value of ${\cal O}(500)$, the system
prefers to have a collection of 5-gons and 7-gons arranged in the
form of a linear chain of alternating 5-gon--7-gon sequence
(``scar''). For even larger values of $N$, the preferred form is
what they name ``pentagonal buttons''. These are configurations of
two nested circles, with five 5-gons placed on the outer circle,
five 7-gons on the inner circle, and finally a 5-gon in the common
center. As $N$ further increases, the defect system formes more
intricate patterns. A $C_3$--symmetric configuration of defects is
noted.

In this paper we would like to give qualitative topological
arguments suggesting why these different defect configurations --
such as pentagonal buttons or $C_3$ symmetric configurations,
discovered by numerical minimization of $E_C$ -- can appear. We
shall do so by recalling in Section II how the topological methods
of Morse \cite{morse} apply to second order phase transitions in
crystals (as pioneered by Michel, see e.g. \cite{Michel}) paving
the way to the application to the Thomson problem we shall deal
with in Section III. Section IV is devoted our conclusions.

%%%
\section{Symmetry breaking, Morse theory and selection rules}
%%%

A crystal has a density function $\rho (\vec{x})$ invariant under
one of the finite crystallographic groups, say $G$
\begin{equation}\label{rho}
    \rho ( g \vec{x} ) = \rho (\vec{x}) \quad {\rm with} \quad g
    \in G \;,
\end{equation}
where $\rho (\vec {x}) dV$ is the probability to find an atom of
the crystal in the volume $dV$. Not all finite groups acting on
$3$ spatial dimensions, $\vec{x} = (x_1, x_2, x_3)$, are
crystallographic groups, i.e. actual symmetries of crystals in
nature. For instance: $O_h$, the {\it octahedral} group, the
largest symmetry group of the cube, is a crystallographic group,
while $I$, the {\it icosahedral} group, largely used in
\cite{Bowick1} for the effective theory of $E_C$, is not.

Crystals undergo second order phase transitions. The theory of
such transitions has been established by Landau \cite{landau}. The
density function $\rho$ changes smoothly from one phase to the
other, while the symmetry group $G$ suddenly changes to a subgroup
$H \subset G$. The density function can then be decomposed as
\begin{equation}\label{deltarho}
    \rho (\vec{x}) + \delta \rho (\vec{x})
\end{equation}
where $\rho$ is $G$-symmetric, while $\delta \rho$ is
$H$-symmetric, and, at temperatures below the critical value $T <
T_c$, $\delta \rho = 0$, while for $T > T_c$ is $\delta \rho \neq
0$. As the function changes continuously during the transition,
$\delta \rho$ is small near $T_c$.

These transitions are (proto-)typical examples of spontaneous
symmetry breaking: The thermodynamic potential $\Phi ( \rho, T,
P)$ is always $G$-invariant\footnote{This is so because it cannot
depend on coordinates, hence, in particular, is invariant under
$G$.}. For $T < T_c$ the vacuum is $G$-invariant, realizing an
explicitly symmetric phase; for $T > T_c$ the vacuum is
$H$-invariant, realizing a spontaneously broken symmetric phase.
By {\it vacuum} here we mean the $\rho$ configuration that
minimizes the functional $\Phi (\rho)$ for the given values of $T$
and $P$.

Thus, given a $G$-symmetric crystal ($\rho$ and $\Phi$) one can
select the sub-group $H$ into which the crystal will make the
transition, among the allowed subgroups of $G$, by finding the
minimum of $\Phi$ with respect to $\rho$. Using physical
reasonings Landau assumed that, near $T_c$, $\Phi$ is a polynomial
in $\rho$ of three terms of order zero, two and four (the Landau
polynomial).

As $\rho$ is a $G$-invariant function, it can be expanded in the
basis of the functions $\{ \phi_i (\vec{x}) \}_{i = 1, ..., {\rm
ord}(G)} $, where ${\rm ord} (G)$ is the number of elements
(order) of the finite group $G$, and
\begin{equation}\label{ordGrep}
\phi_i (\vec{x}) = \sum_{j = 1}^{{\rm ord} (G)} D_{i j} [g] \phi_j
(\vec{x}) \quad i = 1, ..., {\rm ord}(G) \;,
\end{equation}
for any $g \in G$. The matrices $\{ D_{i j} \}$ realize a
reducible representation of $G$. Being $\rho$ real, the
representation is necessarily orthogonal.

In particular
\begin{equation}\label{delta}
    \delta \rho (\vec{x}) = \sum_{i = 1}^{{\rm ord} (G)} \eta_i
    \phi_i (\vec{x})
\end{equation}
and
\begin{equation}\label{deltaprimo}
    \delta \rho' (\vec{x}) = \delta \rho (g \vec{x}) = \sum_{i = 1}^{{\rm ord} (G)} {\eta'}_i
    \phi_i (\vec{x})
\end{equation}
with ${\eta'}_i = \sum_j D_{i j} [g] \eta_j$. The $\eta_i$s are
the {\it order parameters} of the phase transition ($\eta_i = 0$
at $T < T_c$, $\eta_i \neq 0$ at $T > T_c$), hence the
thermodynamic potential will be minimized with respect to these
quantities $\Phi (\vec{\eta})$. The reducible set $\{ \eta_i \}$
can be decomposed into irreducible sub-sets $\{ \eta_i^a \}$ where
\begin{equation}\label{etairrep}
    \eta_i^a = \sum_{b = 1}^n d_{a b} [g] \eta_i^b \quad a = 1,
    ..., n \;,
\end{equation}
the $d_{a b} [g]$s being the $n \times n$ irreducible
matrix-blocks in the ${\rm ord} (G) \times {\rm ord} (G)$ matrices
$D_{i j} [g]$ and, supposing there are $m$ such irreducible
representations of dimension $n_1, n_2, ..., n_m$, we have
\begin{equation}\label{ordG}
    n_1^2 + n_2^2 + ... + n_m^2 = {\rm ord} (G) \;.
\end{equation}
Note that $\{ \eta_i^a \}$ (or the corresponding $\{ \phi_i^a \}$)
is not a complete set. Nonetheless, it is only one such sets that
is needed to expand $\delta \rho$ near $T_c$ \cite{landau}. Thus,
the problem of finding which way the $G$-symmetry of the crystal
is spontaneously broken down to the smaller symmetry of one of its
subgroups -- say $H$ -- boils down to the minimization of the real
function(al)
\begin{equation}\label{Phicompact}
    \Phi : \vec{\eta} \in S^{n} \to \Phi (\vec{\eta})
    \in R
\end{equation}
where $n$ is one of the $m$ values in (\ref{ordG}), and we added
to $R^{n}$ the point at infinity that, for stability, has to be
included as a maximum, so that $R^n + \{ \infty \} = S^n$ . In
what follows we shall take $n= 3$.

One way of solving this problem is, of course, to explicitly know
the coefficients of the Landau polynomial that depend upon the
details of the model. Many things, though, can be said about the
critical points (minima, maxima, saddle points) of a real function
like $\Phi$ in (\ref{Phicompact}), {\it without knowing its
explicit expression}, but rather exploiting the constraints
associated with the topology of $S^n$, as shown in the
mathematical work of Morse \cite{morse} (for a physically
intuitive introduction see, e.g., \cite{Sen}; for extensive
applications to phase transitions in crystals see, e.g.,
\cite{Michel}).

In a nutshell, Morse proved that, when a function is like $\Phi$,
i.e. smooth, real and defined over a compact differentiable
manifold (as said, we shall focus on the case of $S^3$), then the
following constraints hold
\begin{eqnarray}
c_0 & \ge & b_0 \label{morse1} \\
c_1 - c_0 & \ge &  b_1 - b_0 \\
c_2 - c_1 + c_0 & \ge & b_2 - b_1 + b_0 \\
c_3 - c_2 + c_1 - c_0 & = & b_3 - b_2 + b_1 - b_0 \label{morse2}
\end{eqnarray}
where $c_\ell$ is the number of critical points of $\Phi$ with
index\footnote{In the neighborhood of $\vec{x}_0$, a result of
Morse theory states that the function $f$ has a local description
of the form
\[
  f (\vec{x}) = f (\vec{x}_0) - y_1^2 - y_2^2 - \cdots - y_\ell^2 +
  y_{\ell + 1}^2 + \cdots + y_{{\rm dim} M}^2 \;,
\]
where $y_i = (\vec{x} - \vec{x}_0)_i$ is the $i^{\rm th}$
coordinate of the vector $\vec{x} - \vec{x}_0$, and ${\rm dim} M$
is the dimension of the manifold $M$. The index $\ell$ of the
critical point $\vec{x}_0$ refers to the number of negative terms
present in the local description of $f(\vec{x})$. Then $\ell = 0$
means that no negative $y^2$ term is present in the local
representation for $f$, so that the critical point is a local {\it
minimum}. Similarly if $\ell = {\rm dim} M$, then it means that
$f$ decreases in value in all the ${\rm dim} M$ directions of the
manifold, hence this critical point is a {\it maximum}.} $\ell$
($\ell = 0$ are the minima, $\ell = 3$ the maxima, $0 < \ell < 3$
the saddle points), and the $b_\ell$s are the Betti numbers of
$S^3$. Recall that
\begin{equation}\label{betti}
    b_0 (S^3) = 1 = b_3 (S^3) \;,
\end{equation}
while the other Betti numbers of $S^3$ are all zero. This is a
powerful result: without knowing the actual form of the function,
the topology of the manifold tells us that for each $\ell$ there
are at least $b_\ell$ critical points with index $\ell$
\begin{equation}\label{cgeb}
    c_\ell \ge b_\ell \;.
\end{equation}

For the case in point we know that $\Phi$ has at least one minimum
($b_0 = 1$) and at least one maximum ($b_3 = 1$). Furthermore,
denoting by $\vec{\eta}^{(\ell)} \in S^3$ the critical point of
$\Phi$ with index $\ell$, and by $H^{(\ell)}$ the subgroup of $G$
that leaves it invariant,
\begin{equation}\label{hinv}
    \eta_a^{(\ell)} = \sum_{b = 1}^3 d_{a b} [h] \eta_b^{(\ell)} \quad h
    \in H^{(\ell)} \;,
\end{equation}
the number of elements (order) of $H^{(\ell)}$ is simply related
to $c_\ell$
\begin{equation}\label{ordHk}
    {\rm ord} (H^{(\ell)}) = {\rm ord} (G) / c_\ell \;.
\end{equation}
It is a result of the theory of finite groups that ${\rm ord}
(H^{(\ell)})$ as written in (\ref{ordHk}) is indeed an integer,
the proof being based on the coset decomposition of $G$ with
respect to $H^{(\ell)}$ \cite{hamermesh}, \cite{Sen}.

The subgroup we are looking for is $H^{(0)}$, the symmetry of the
vacuum configuration $\vec{\eta}^{(0)}$ and new symmetry $H$ of
the crystal beyond $T_c$. The constraints
(\ref{morse1})-(\ref{morse2}) and (\ref{ordHk}), the requirement
to have {\it at least} two maxima (at $\vec{\eta}^{(3)} = \infty$
and at $\vec{\eta}^{(3)} = 0$), i.e. $c_3 \ge 2$, and the fact
that the number of \textit{real} solutions of $d \Phi / d
\vec{\eta} = 0$ (with $\vec{\eta} \in S^3$ and $\Phi$ a fourth
order polynomial) is bounded to be
\begin{equation}\label{boundrealsol}
    c_0 + c_1 + c_2 + c_3 \le 3^3 +1 = 28 \;,
\end{equation}
prove to be enough to identify the permitted $H^{(0)}$s! Which one
is the actual $H$ of the crystal after the spontaneous breaking of
symmetry $G \hookrightarrow H$, of course, depends upon the
particular physical situation under investigation.

Examples are crystalline structures, such as the alloy Cu-Zn, that
exhibit the octahedral symmetry, whose group $O_h$ has order 48.
The subgroups of $O_h$ allowed by the requirement to be
crystallographic groups, have order 8, 6, 4 and 2. Carrying on the
analysis outlined above furnishes: ${\rm ord} (H^{(0)}) = 6$,
which corresponds to $C_{3 v}$, and ${\rm ord} (H^{(0)}) = 8$,
which corresponds to $C_{4 v}$ (recall that ${\rm ord} (C_n) = n$
and  ${\rm ord} (C_{n v}) = 2 n$). This means that $C_{3 v}$ and
$C_{4 v}$ are {\it selected} as the only possible symmetries of a
crystal, originally $O_h$-symmetric, undergoing a second order
phase transition. More on this can be found in \cite{Michel} and
in \cite{Sen}.

%%%
\section{The Thomson Problem in the Icosahedral Approximation}
%%%

We shall now extensively rely on the model built up in
\cite{Bowick1}, which deals with a system of topological defects
on the surface of a spherical elastic material. Our considerations
will be qualitative in nature, hence we shall skip many of the
details of that analysis and will focus only on the topological
and group-theoretical aspects of the problem. This way we shall
provide an explanation of (some) of the empirical results of that
work with the hope that our analysis could provide useful insight
for that problem and, more generally, for the Thomson problem.

The authors of \cite{Bowick1} deal with a sphere, originally tiled
with an arbitrary number of hexagons and exactly twelve pentagons.
The topology of the sphere requires that the defects, emerging as
\textit{disclinations} in the underlying structure of hexagons,
must be: first the initial 12 pentagons, then the pairs
pentagons-heptagons. On this we commented in the Introduction. For
interactions among the defects that are strong compared to the
interactions among the particles forming the material, as
collectively described by the Young modulus, (what in
\cite{Bowick1} is identified as the ``large core'' energy regime),
the twelve pentagons are the only defects in the system and are at
the vertices of an \textit{icosahedron}, because this
configuration minimizes the energy\footnote{We shall still call
the energy $E_C$. This is a clear abuse of notation because the
energy here is quite more complicated than the expression given in
(\ref{ethom}). Nonetheless, for our considerations the actual form
of the function is not important.} $E_C$. In the ``small core''
energy regimes new defects are energetically allowed as 12-plets
of pentagon-heptagon pairs, and again they prefer to arrange
themselves at the vertices of icosahedra. The patterns these
additional defects describe on the surface of the sphere is our
concern.

To try to understand that, we shall apply the analysis of the
previous Section by considering the energy as a functional of the
density function. Thus we shall regard $E_C$ as a map
\begin{equation}\label{ECMorse}
E_C: \vec{\eta} \in S^n \to R \;,
\end{equation}
where, as explained, $n$ is related to the symmetry group, the
icosahedral group $I$ in this case, the point at infinity is added
for stability as a maximum, and the notation is the same as
before. In the large core energy regime $E_C$ and its
minimum\footnote{Recall that (in the language of the Morse
inequalities) $c_0 = {\rm ord}(I) / {\rm ord}(I) = 1$.} (vacuum)
are $I$-symmetric. In the small core energy regimes we shall make
the approximation that $E_C$ is still $I$-symmetric\footnote{That
the $I$-symmetry of $E_C$ in the small core energy regimes is a
good approximation is a result we borrow from \cite{Bowick1},
where it is shown that indeed the energy is not severely modified
by the introduction of these new defects.}, but the vacua will
only be invariant under a subgroup of $I$. This way we are dealing
with the spontaneous breaking of the icosahedral symmetry into
smaller symmetries we want to determine.

The icosahedron is a polyhedron with twelve vertices, twenty faces
and thirty edges. Its symmetry group $I$ has 60 elements arranged
into 5 conjugacy classes (the number of elements in the class is
in parentheses)\cite{hamermesh}: Identity (1), $C_2$ (15), $C_3$
(20), $C_5$ (12), $C_5^2$ (12). The subgroups of $I$ are $C_2$,
$C_3$, $C_5$ (of order 2, 3, 5, respectively) and those obtained
by the coset decomposition of $I$ with respect to them. The list
of the \textit{orders} of the subgroups is
\begin{equation}\label{subgroupsorder}
    \{ 2, 3, 5, 12, 20, 30 \} \;.
\end{equation}
Of course, by definition of polyhedron, the vertices of the
icosahedron lie on the surface of a sphere. Nonetheless, we want
to stress this here because, together with the topological
constraints on the tiling and the choice of the irreducible
representation of $I$ relevant for the problem in point (i.e. the
\textit{three-dimensional} one), it makes clear the role of the
two-sphere in configuration space. These comments are also in
order because the topology of a \textit{different} sphere, the
$S^n = S^3$ in (\ref{ECMorse}), will now play an important role.

Near the phase transition we suppose that $E_C$ is a polynomial of
order four, just as for the thermodynamic potential $\Phi$ of
crystals. Hence the number of real solutions of ${E'}_C = 0$ is
bounded from above
\begin{equation}\label{boundrealsolicosahedral}
    c_0 + c_1 + c_2 + c_3 \le 3^3 + 1 = 28 \;,
\end{equation}
where the point at infinity is also added.

We can now use the relevant Morse inequalities for the function in
(\ref{ECMorse}) with $n=3$
\begin{eqnarray}
c_0 & \ge & 1 \label{morseico1} \\
c_0 & \le &  1 + c_1 \\
c_0 + c_2 & \ge & 1 + c_1 \\
c_0 + c_2 & = & c_1 + c_3 \;. \label{morseico2}
\end{eqnarray}
These, the request that $c_3 \ge 2$ and the constraint
(\ref{boundrealsolicosahedral}) are what we need to identify the
order of the allowed subgroups $H^{0}$. We just have to insert the
given order from the list (\ref{subgroupsorder}) into
\begin{equation}\label{c0ico}
    c_0 = 60 / {\rm ord} (H^{0}) \;.
\end{equation}
The resulting list of allowed subgroups consistent with the Morse
inequalities and with the constraint
(\ref{boundrealsolicosahedral}) is
\begin{equation}\label{subgroupsorderselected}
    \{ 5, 12, 20, 30 \} \;.
\end{equation}
Let us analyze the case ord$(H^{0})$ = 12. In this case the
$I$-symmetry has been broken in such a way that there are 5
distinct minima ($c_0 = 5$) connected by a $C_5$ transformation
(recall that 12 is the order of the coset of $I$ with respect to
$C_5$). Each minimum corresponds to a new icosahedron of defects,
thus, according to this argument, there would be 5 new icosahedra.
But, as shown, the new defects cannot appear unless they are in
pairs of 5-gons--7-gons, thus, being 5 odd, we {\it must} have 10
new icosahedra: 5 icosahedra of pentagons next to 5 icosahedra of
heptagons. These are the pentagonal buttons found in
\cite{Bowick1}. It would be interesting to see how the residual
symmetry of order twelve is related to the way these buttons
arrange themselves on the sphere which, according to
\cite{Bowick1}, is at certain vertices of the rhombic
tricontahedron.

Similarly the case ord$(H^{0})$ = 20 refers to 3 distinct minima
for the new symmetry, connected by a $C_3$ transformation. By
means of the same constraint to have 5-gons--7-gons pairs, we
would expect to see 6 new icosahedra at the time, while the
numerical results show 7 such clusters. This discrepancy could be
due to the approximations we are introducing, either for the
invariance of the energy or of the order of the polynomial near
the phase transition.

We find also the cases: ord$(H^{0})$ = 30, which refers to 2
distinct minima connected by a $C_2$ transformation, and the case
ord$(H^{0})$ = 5 which means $H = C_5$, in the notation of the
previous Section.

%%%
\section{Conclusions}
%%%

We applied the method of Morse theory and second order phase
transition in crystals to the Thomson problem in the icosahedral
approximation of reference \cite{Bowick1}. We are able to describe
the pentagonal buttons and the $C_3$ symmetry found there,
although this last one only to an approximated extent. We also
notice residual symmetries of orders not seen in that numerical
work. This does not necessarily means that these new symmetries
are to be found exactly in form given, because the analysis we
made might require further care in order to be applied to other
cases.

Our results are qualitative and preliminary. For instance, within
this approach we are not able to address the important issue of
the various energy thresholds involved with the phase transitions.
Thus we cannot say {\it when} the transitions happen. Nonetheless,
we can address the question {\it why} these transitions occur and
in such a general fashion that it is suitable, in principle, for a
wide range of applications. Hence, we believe, the results
presented here are interesting especially in the view of
stimulating the discussion for a deeper understanding of the
Thomson problem in general.

We also notice that, having clarified the topological origin of
the pentagonal buttons and $C_3$-symmetric configurations, it is
reasonable to argue that these configurations share some features
of \textit{solitons}, in the sense that they cannot be undone by a
continuous transformation.

\acknowledgments

\noindent A.I. thanks the School of Theoretical Physics of the
Dublin Institute for Advanced Studies and S.S. thanks the Physics
Department ``E.R. Caianiello'' of Salerno University, for their
kind hospitality while some of this work was completed.

\end{document}